\documentstyle[12pt]{article}
\begin{document}
\bibliographystyle{unsrt}

\begin{center}
{\large \bf ELECTROWEAK PROPERTIES OF LIGHT MESONS  IN RELATIVISTIC 
HAMILTONIAN DYNAMICS}\\

\vbox {\vspace{2mm}}
A.F.Krutov\\
Department of Physics, Samara State University, Samara, 443011,Russia\\[3mm]
V.E.Troitsky\\
Moscow State University\\D.V.Skobeltsyn Institute of Nuclear
Physics, Moscow, 119899,Russia\\

\end{center}
\begin{abstract}
The calculation of lepton decay constants and electromagnetic form factors
of pion and kaon is performed in the framework of relativistic
hamiltonian dynamics. The different model wave function is used. Wave
function parameters are fixed from fit of mean square radius of meson.
The internal quark structure is taken into account by electromagnetic
quark form factor and quark anomalous magnetic moment. The pion and kaon
form factors depend only weakly on model wave function. Strong dependence
of these values on quark anomalous magnetic moment is obtained in the transfer
momentum region of CEBAF experiments.
\end{abstract}

{\large \bf Introduction}

At the present time the interest  for the description of light mesons 
electroweak structure surfaced \cite{ChC88}--\cite{CaG95}. 
This interest is initiated by future CEBAF
experiments on measurement of elastic pion and kaon form factors at
momentum transfer less then 3 (GeV/c)$^2$ (E--93--021 and E--93--018).

These experiments can give the possibility of choosing between different
approaches which are distinguished by starting position as well as the 
results.

Our report is devoted to the description of electroweak properties of light
mesons as two-quark systems in the framework of so called relativistic
Hamiltonian dynamics (RHD) or Dirac's dynamics (see, e.g. \cite{KeP91}).
At the present time this approach is used widely for relativistic description
of composite systems in particle and nuclear physics: 
\cite{ChC88}--\cite{KrT93}, \cite{CaG94}--\cite{CaG95}.

RHD can be formulated in different forms \cite{KeP91},\cite{Lev93}. 
The most popular at the present time is
light-front form which is used for description of the electroweak properties
of pion and kaon in many works ( \cite{ChC88},
\cite{CaG94}--\cite{CaG95}).

In our work the other form of relativistic is used, i.e. instant
form of dynamics (see e.g. \cite{BaK95}). This approach was successfully
applied to description of electromagnetic properties of light mesons
\cite{KrT93}, \cite{BaK96}.

In this report the results of calculation of lepton decay constants and
electromagnetic form factors for pion and kaon are presented. It turns out
that the satisfactory description of lepton decay constants and mean square 
radius (MSR) of light mesons is impossible simultaneously without taking into
account internal quark structure. Internal quark structure is incorporated
by electromagnetic quark form factor and anomalous quark magnetic moments.
Parameters of wave functions are fixed by condition of description of MSR in 
the limits of experimental errors.

The pion and kaon
form factors depend only weakly on model wave function. On the other hand
strong dependence
of form factors on quark anomalous magnetic moment is taken place in the 
transfer momentum region of CEBAF experiments.

{\large \bf Lepton decay constant of mesons}

Lepton decay constant of pion or kaon is determined by following matrix 
element \cite{Jaus91}:
\begin{equation}
<0|j^\mu|\,p_c\,> = if_c\,p_c\,^\mu\frac{1}{(2\pi)^{3/2}}\>.
\label{j=f_c}
\end{equation}
$f_c$ -- lepton decay constant, $p_c$ -- 4-momentum of meson.

In RHD the Hilbert space of composite particle states is the tensor
product of single particle Hilbert spaces:
${\cal H}_{q\bar Q} \equiv  {\cal H}_{q} \otimes
{\cal H}_{\bar Q}$
and the state vector in RHD is a superposition of two-particle
states. 
As a basis in
${\cal H}_{q\bar Q}$ one can choose the following set of vectors:
\begin{equation} |\,\vec P,\>\sqrt
{s},\>J,\>l,\>S,\>m_J\,>\>,
\label{bas-cm}
\end{equation}
with $P_\mu = (p_1 +p_2)_\mu$,
$P^2_\mu = s$, $\sqrt {s}$ ---
the invariant mass of two-particle system , $l$
--- the angular moment in the center-of-mass frame,
$S$ --- total spin,
$J$ --- total angular momentum, $m_J$ ---
projection of total angular momentum.

After decomposition of matrix element (\ref{j=f_c}) in basis (\ref{bas-cm})
we obtain the following expression for lepton decay constant: 
\begin{equation}
\int\,d\sqrt{s}\,
G_0(s)
\varphi(k) 
= f_c\>,
\label{int G_0(s)=f_c}
\end{equation}
%\begin{equation}
$$
G_0(s) = \frac{n_c}{2\sqrt{2}\,\pi\,P_0}\sqrt{(p_{10}+M_1)(p_{20}+M_2)}\,
\left[1 - \frac{k^2}{(p_{10}+M_1)(p_{20}+M_2)}\right]\>.%\label{G0}
%\end{equation}
$$
$$
p_{i0} = \sqrt{k^2 + M_i^2}\>,\quad i=1\>,\>2
$$
$n_c$ -- number of quark colors.
For $\varphi (k)$ one can use an any phenomenological
wave function which will discussed below.
For the case of the pion it is necessary to substitute in all expressions
$M_1=M_2=M$.

Details of calculations can be found in work \cite{Kr97}.

{\large \bf Electromagnetic form factors of mesons}

Electromagnetic form factor of pion or kaon is determined by electromagnetic
matrix element:
\begin{equation}
<p_c|\,j_\mu\,|{p'}_c>=(p_c+{p'}_c)_\mu\,F_c (Q^2),
\label{j=F_c}
\end{equation}
$F_c(Q^2)$ -- electromagnetic form factor of meson.

Using basis (\ref{bas-cm}) we obtain following expression for meson form 
factor in the instant dynamics:
\begin{equation}
F_c (Q^2)=\int d\sqrt s\ d\sqrt{s'}\ \varphi (k)\,g_0(s,Q^2,s')\,
\varphi (k').
\label{F_c}
\end{equation}
$g_0(s,Q^2,s')$ -- so called free two particle form factor. Free two particle
form factor was calculated in ref.\cite{BaK96}. To save space we shall not
present corresponding expression here. $\varphi (k)$ has the same meaning as
in eq. (\ref{int G_0(s)=f_c}).

Details of calculation can be found in work \cite{BaK96}.

For the calculation of MSR we used corresponding definition:
\begin{equation}
<r^2>=\left.-\,6\,dF_\pi(Q^2)/dQ^2 \right|_{Q^2=0}% = <r^2>_{exp}
\label{rad}
\end{equation}

\pagebreak

{\large \bf Wave function of mesons}

Wave functions in RHD are given as eigenfunction of the complete 
commuting operators set. For instant form dynamics this set is following:
${\hat M}^2_I,\>{\hat J}^2,\>\hat J_3,\>\hat
{\vec P}$. ${\hat M}^2_I$ -- the mass square operator of the composite system,
${\hat J}^2$ -- the total angular momentum square operator, 
$\hat J_3$ -- operator of third component of total angular momentum,
$\hat{\vec P}$ --  the operator of total 3-momentum.

To obtain the wave function of the system one needs to diagonalize
this set. In the case of instant form dynamics the operators ${\hat J}^2,
\>\hat J_3,\>\hat {\vec P}$ coincide with the
appropriate operators of the two-particle system without interaction and one
can construct the basis (\ref{bas-cm}) in which these three operators are 
diagonal. While
working in this basis to obtain the wave function one needs to diagonalize
${\hat M}^2_I$. Corresponding eigenvalue problem is reduced to Schr\"odinger--
like equation \cite{KeP91},\cite{BaK95}.
Corresponding wave function is the following:
$$
<\vec P\,',\,\sqrt
{s'},\,J',\, l',\,S',\,m_J'|\,p_c >\> =$$ 
\begin{equation}
=N_C\,\delta (\vec P\,' -
\vec p_c)\delta _{JJ'}\delta _{m_Jm_J'}
\delta _{ll'}\delta _{SS'}\,\varphi
(k)\>,\label{wf}
\end{equation}
$$
k = \frac{[s^2 - 2s(M_1^2 + M_2^2) + \eta^2]^{1/2}}{2\sqrt{s}}\>,\quad \eta =
M_1^2 - M_2^2\>, 
$$
$$
N_C = \sqrt{2p_{c0}}\sqrt{\frac{N_{C-G}}{4\,k}}\>,\quad
N_{C-G} = \frac{(2P_0)^2}{8\,k\,\sqrt{s}}\>,
$$
In the case of pion and kaon $J = l = S = m_J = 0$.

For $\varphi (k)$ one can use an any phenomenological
wave function, normalized using the relativistic density of
states:
\begin{equation}
\varphi(k(s)) =\sqrt{\sqrt{s}(1 - \eta^2/s^2)}\,u(k)\,k\>,\label{phi(s)}
\end{equation}         
Condition of normalization is
\begin{equation}
n_c\,\int\,u^2(k)\,k^2\,dk = 1\>.\label{norm}
\end{equation}
$n_c = 3$ -- number quark colors.

$u(k)$ - is nonrelativistic phenomenological wave function.

The
following wave functions were utilized in our work:

1. A gaussian or harmonic oscillator (HO)
wave function (see e.g.  \cite{ChC88})
\begin{equation}
u(k)= N_{HO}\,
\hbox{exp}\left(-{k^2}/{2b^2}\right),
\label{HO-wf}
\end{equation}

2. A power-law (PL) wave function (see e.g. \cite{CaG94},\cite{Sch94b}) 
\begin{equation}
u(k) =N_{PL}\,{(k^2/b^2 +
1)^{-n}}\>,\quad n = 2\>,3\>.  \label{PL-wf}
\end{equation}

3. The wave function with linear confinement from Ref.\cite{Tez91}:
\begin{equation}
u(r) = N_T \,\exp(-\alpha r^{3/2} - \beta r)\>,\quad \alpha =
\frac{2}{3}\sqrt{2\,M_r\,a}\>,\quad \beta = M_r\,b\>.
\label{Tez91-wf}
\end{equation}
$a\>,b$ -- parameters of linear and Coulomb parts of potential
respectively, $M_r$ -- reduced mass of two-particle systems.

{\large \bf Calculations and results}

In our calculations we have fixed the masses of u- and d-quarks to be
equal to 0.25 GeV, and that of s-quark from the approximate relation
$M_s/M_u \approx$ 1.4. These values are usually used in relativistic
calculations.

We have used $b = $ 0.7867 for the
model (\ref{Tez91-wf}) which means that the strong coupling constant is
equal to 0.59 in the light meson mass scale.

We have imposed the conditions of description in the experimental region
errors of lepton decay constants (\ref{int G_0(s)=f_c}) and MSR of pion and
kaon (\ref{rad}) \cite{Amen84}--\cite{PDG}:
$$
<r^2_\pi> =\hbox{0.432} \pm \hbox{0.016 fm}^{-2}\>,\quad
<r^2_K> =\hbox{0.34} \pm \hbox{0.05 fm}^{-2}\>,
$$
$$
f_\pi =\hbox{131.7} \pm \hbox{0.2 MeV}\>,\quad
f_K =\hbox{160.6} \pm \hbox{1.4 MeV}\>,
$$

The conditions of satisfactory description of MSR and lepton decay constant at 
the
same parameters can not be fulfilled in the model of point-like quark. For
the satisfactory description of MSR and lepton decay constant it is 
necessary to take into account the internal quark structure.

The internal quark structure can be described by introduction of 
electromagnetic quark
form factor and anomalous quark magnetic moment. Electromagnetic quark
structure was chosen in the form:
\begin{equation}
f_q(Q^2) = \frac{1}{1 + \ln(1+ <r^2_q>Q^2/6)}\>.\label{f_qour}
\end{equation}
$<r^2_q>$ -- is MSR of the quark. The choice of form (\ref{f_qour}) does not
violate the asymptotic of meson form factor which takes place in our approach
at $Q^2\>\to\>\infty\>,\> M_q\>\to\>$0:
\begin{equation}
F_c(Q^2) \quad \sim\quad Q^{-2}\>.\label{Q-inf}
\end{equation}
MSR of the quark was calculated from relation \cite{CaG95}:
\begin{equation}
<r_q^2> = \alpha/M_q^2\quad \hbox{at}\quad
\alpha\>\simeq\>\hbox{0.3}\>.\label{r_q}
\end{equation}
Anomalous quark magnetic moments can be found from Gerasimov's sum rules
\cite{Ger93}:
\begin{equation}
\frac{\mu_u}{\mu_d} = \frac{q_u + \kappa_u}{q_d + \kappa_d} = -1.77\>,\quad
\left(\frac{\mu_u/\mu_d + \mu_s/\mu_d}{1 + \mu_s/\mu_d}\right)^2 = 
0.42\pm 0.03\>,\label{Ger93}
\end{equation}
here $\mu_a$ -- magnetic moments of $u,\>d,\>s$ -- quarks in natural units,
$q_a$ -- charge of quark, $\kappa_a$ -- anomalous quark magnetic moment.

For the calculation we have varied sum of anomalous quark magnetic moment 
$\kappa_u + \kappa_{\bar d}$ from 0.09 to --0.09 in natural units.
Lepton decay constants are calculated in assumption of point-like quarks.

The results of our calculation for the pion are presented on the figure. 
 Curves 1--3 give the results of calculations with following
anomalous quark magnetic moments:
$\kappa_u = \hbox{-0.085}\>,\> \kappa_d = \hbox{0.005}\>,\quad \kappa_s =
\hbox{0.11}$ ($\kappa_u + \kappa_{\bar d}$ = -0.09). 1-- model (\ref{HO-wf})
$b$=0.325 GeV,$f_\pi$ = 131.6 MeV,\\ 
2-- model (\ref{PL-wf}), $n$=2, $b$=0.396
GeV, $f_\pi$ = 144.3 MeV, 3-- model (\ref{PL-wf}), $n$=3, $b$=0.571,
$f_\pi$ = 135.8 MeV and model (\ref{Tez91-wf}), $b$=0.7867, 
$a$=0.083 GeV$^2$,
$f_\pi$ = 136.1 MeV. Results of models (\ref{PL-wf}) with $n$=3 and 
 (\ref{Tez91-wf}) are practically coincide. Curves 4 gives the results of
model (\ref{HO-wf}) with $\kappa_u = \hbox{0.03}\>,\quad \kappa_d = 
\hbox{-0.06}\>,\quad \kappa_s =
\hbox{0.065}$ ($\kappa_u + \kappa_{\bar d}$ = 0.09), 
$b$=0.278 GeV,$f_\pi$ = 121.0 MeV.

For all of the models we listed the calculated values of pion lepton decay 
constant.

So one can see comparatively weak model dependence of electromagnetic 
pion form factor in region of CEBAF experiments (let us compare 
the curves 1--3
at $Q^2\><\>$3 (GeV/c)$^2$). On the other hand,
pion electromagnetic form factor depend strongly on the anomalous
quark magnetic moments at condition of correct description of pion MSR
(\ref{rad}) (let us compare the curves 1 and 4 at $Q^2\><\>$3 (GeV/c)$^2$). 
Experimental points are taken from ref. \cite{Amen84},\cite{Beb78}.

Analogical results take place for kaon.

Based on our results it is possible to do the following conclusions.

1. Most likely the CEBAF experiments on measurements of $\pi$-- and $K$--
mesons form factors at $Q^2\>\leq\>$3 (GeV/c)$^2$ will not allow for the 
possibility
to extract the best model of quark interaction in view of weak dependence
on choice of the wave function. 

2. It seems reasonable to say that CEBAF
experiments will give the possibility to estimate the manifestation of 
internal quark structure.\\[0.35cm]

This work is supported in part by Russian Foundation for Basic Research 
(Grant no. 96--02--17288).\\[13cm]

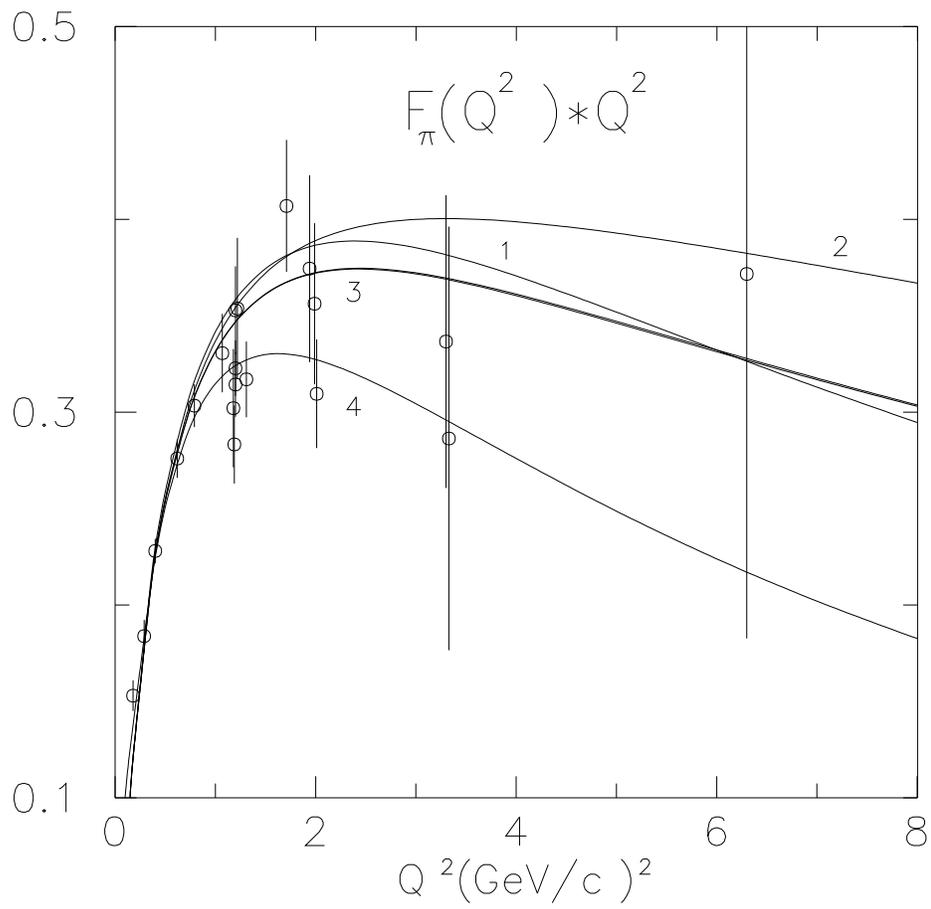
\end{document}